\documentclass[aps,prb,twocolumn,superscriptaddress,showpacs]{revtex4}

\usepackage{longtable}
\usepackage{graphicx}
\usepackage{dcolumn}
\usepackage{bm}
\usepackage{amsmath}
\usepackage[psamsfonts]{amssymb}

\newcommand{\PCCO}{Pr$_{2-x}$Ce$_x$CuO$_4$}
\newcommand{\LSCO}{La$_{2-x}$Sr$_x$CuO$_4$}

\newcommand{\QPT}{quantum phase transition}
\newcommand{\QCP}{quantum critical point}

\newcommand{\NCCO}{Nd$_{2-x}$Ce$_x$CuO$_4$}

\newcommand{\MR}{magnetoresistance}
\newcommand{\nMR}{negative magnetoresistance}

\newcommand{\etal}{\emph{et al.}}

\begin{document}

\title{Anomalous resistivity saturation at low temperatures in Pr$_{2-x}$Ce$_x$CuO$_{4-
\delta}$}

\title{On the resistivity at  low temperatures in electron-doped cuprate superconductors}


\author{S. Finkelman}
\affiliation{Raymond and Beverly Sackler School of Physics and Astronomy, Tel-Aviv
University, Tel Aviv, 69978, Israel}
\author{M. Sachs}
\affiliation{Raymond and Beverly Sackler School of Physics and Astronomy, Tel-Aviv
University, Tel Aviv, 69978, Israel}
\author{J. Paglione}
\affiliation{ center for nanophysics and advanced materials, Physics Department, University of Maryland, College Park,
Maryland 20743, USA}
\author{G. Droulers}
\affiliation{ center for nanophysics and advanced materials, Physics Department, University of Maryland, College Park,
Maryland 20743, USA}\author{P. Bach}
\affiliation{ center for nanophysics and advanced materials, Physics Department, University of Maryland, College Park,
Maryland 20743, USA}
\author{R.L. Greene}
\affiliation{ center for nanophysics and advanced materials, Physics Department, University of Maryland, College Park,
Maryland 20743, USA}
\author{Y. Dagan}
\email[]{yodagan@post.tau.ac.il}
\affiliation{Raymond and Beverly Sackler School of Physics and Astronomy, Tel-Aviv University, Tel Aviv, 69978,
Israel}


\date{\today}

\begin{abstract}
We measured the magnetoresistance as a function of
temperature down to 20mK and magnetic field for a set of underdoped Pr$_{1.88}$Ce$_{0.12}$CuO$_{4-\delta}$ thin films with controlled oxygen content. This allows us to access the edge of
the superconducting dome on the underdoped side. The sheet resistance
increases with increasing oxygen content whereas the superconducting
transition temperature is steadily decreasing down to zero. Upon applying various magnetic fields to suppress superconductivity we found that the sheet resistance increases when the temperature is lowered. It saturates at very low temperatures. These results, along with the magnetoresistance, cannot be described in the context of zero temperature two dimensional superconductor-to-insulator transition nor as a simple Kondo effect due to scattering off spins in the copper-oxide planes. We conjecture that due to the proximity to an antiferromagnetic phase magnetic droplets are induced. This results in negative magnetoresistance and in an upturn in the resistivity .
\end{abstract}


\maketitle

\section {Introduction}

The electron doped (n-doped) cuprates,
(RE$_{2-x}$Ce$_x$CuO$_{4-\delta}$ with RE=Nd, Pr, La, Sm),
superconductors offer a unique system for studying the low
temperatures normal state properties of a high T$_c$ cuprates. In
most of the high T$_c$ cuprates a very high field is needed to quench superconductivity. The normal
state is thus obscured by the occurrence of superconductivity. In
the n-doped cuprates the normal state is accessible at a relatively modest
magnetic fields $(H<10T)$.
\par
Of particular interest is the insulating like behavior at low temperatures.
The derivative of the resistivity becomes negative below a certain temperature $T_m$ for optimum doping $(x=0.15)$ and for the underdoped regions on both the electron and hole doped sides of the phase diagram.\cite{andoboebingerfirst, fournierPRL} Steiner \etal~\cite{SteinerKapitulnik} suggested that the behavior of \LSCO~ at magnetic fields high enough to quench superconductivity is similar to that of thin films of amorphous indium oxide near a magnetic field tuned superconductor to insulator (SIT) phase transition. The scaling behavior suggested by M. P. A. Fisher~\cite{FisherScalin} for SIT was demonstrated for underdoped YBa$_2$Cu$_3$O$_{7-\delta}$.\cite{SITYBCO} In this model a finite and universal value for the critical sheet resistance is predicted at a critical field, identified as the crossing point of the \MR~ isotherms. This crossing point is one of the hallmarks of a field tuned SIT.
\par
Bobroff \etal\cite{Bobroff} showed that introducing nonmagnetic impurities such as Zn or Li in  YBa$_2$Cu$_3$O$_{7-\delta}$ results in a behavior analogous to the Kondo effect observed for dilute alloys with magnetic impurities. This behavior was referred to as "Kondo-like". Rullier-Albenque \etal~ observed similar behavior for electron irradiated optimally-doped and underdoped YBa$_2$Cu$_3$O$_{7-\delta}$ at intense magnetic fields.\cite{Rullier} They suggested that for cuprates without irradiation damage the upturn can be due to the presence of small amounts of disorder. This is in line with a theoretical study finding that a small number of nonmagnetic impurities might induce a Kondo-like behavior at low temperatures when the system is close to an antiferromagnetic transition.\cite{Kontani_nonmagnetic_imps}
\par

\begin{figure}
\resizebox{0.95\columnwidth}{!}{%
  \includegraphics{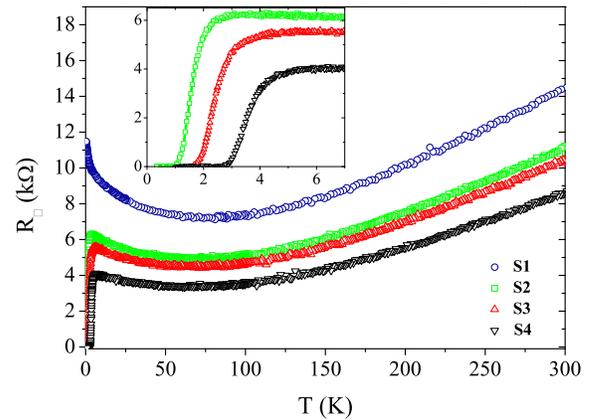}
}

\caption{(color online) Sheet resistance as a function of temperature for the various samples. Inset:zoom on the low temperature region for the superconducting samples (S2-S4).
\label{allrt}}
\end{figure}


In electron-doped cuprates Fournier \etal\cite{fournier2dWL} interpreted the upturn in resistivity,
as well as the \nMR~, as a result of a two dimensional weak
localization. Sekitani \etal\cite{seikitani} suggested that the resistivity upturn and the
\nMR~ are due to scattering off Cu$^{2+}$ Kondo impurities induced
by residual apical oxygen. On the other hand, it has been shown that the doping dependence of the resistivity and Hall coefficient at low
temperatures are characteristic of a system near a \QPT.\cite{daganResistivityPRL} In addition, the \nMR~ is comprised of two contributions: an orbital \MR~ and \MR~ due to spin scattering that vanishes at this \QCP.
At this critical doping $T_m$ goes to zero.\cite{daganMR} This behavior was interpreted in terms of scattering off magnetic droplets existing for the under and optimum doping levels.\cite{daganMR}
A theoretical study by Chen Andersen and Hirschfeld has found that magnetic droplets induced by disorder in underdoped cuprates can lead to an insulating-like behavior and to an upturn in the resistivity.\cite{Hirschfeld_upturns} Gantmakher \etal\cite{RusianSIT} fit the magnetoresistance curves for \NCCO~ with a theoretical formula including both superconducting fluctuations and the Aronov-Altshuler electron-electron interaction correction. Their sample, however, has T$_c\sim12 K$ well above the doping level reported here.
\par
For the electron-doped cuprates, excess oxygen has to be removed in order to obtain superconducting samples. This is usually done in a low oxygen pressure annealing step.\cite{Meisergrowthcond}
There are two main scenarios which explain the sensitivity to oxygen reduction (0.1$\%$ change of oxygen content has a similar effect on T$_c$ to a change of 0.05-0.1 in cerium doping). Higgins \etal\cite{higgins:104510} found that in addition to its doping effect, excess oxygen results in strong scattering. This affects the residual resistivity. In addition T$_c$ is strongly suppressed by this scattering. They concluded that upon reduction, oxygen mainly comes out from the impurity apical sites in the T' structure. By contrast, Raman and infrared-transmission studies suggest that oxygen is removed mainly from the CuO$_2$ planes. This results in destruction of the long-range antiferromagnetism and in the appearance of superconductivity.\cite{RiouOxygen1, RichardOxygeninreduction}The doping at which the resistivity upturn and maximum resistivity appear are somewhat different for various kinds of electron-doped cuprates.\cite{KuiLSCO, Krockenbergerphasediagram}

In this paper we study the resistivity upturn at the edge of the superconducting dome of \PCCO. To tune T$_c$ down to zero we use a cerium doping of x=0.12 together with a small amount of excess oxygen. We examine the \MR~ and resistivity at the transition from the superconducting phase to the nonsuperconducting one. We find that they do not fit to the simple SIT scenario nor to the standard Kondo one. We interpret our data in terms of magnetic droplets resulting in strong spin scattering.

\section {Sample preparation and measurements}

We used pulsed laser deposition (PLD) to grow a set of Pr$_{1.88}$Ce$_{0.12}$CuO$_{4-\delta}$ films with a thickness of 1600\AA~on LaSrGaO$_4$ substrates. The films were grown using the conditions described elsewhere\cite{diamant_singleE}, but annealed in situ at different pressures of N$_2$O: 200 (S1), 9 (S2), 5 (S3), and 0.5 miliTorr (S4). Annealing time was long enough to establish diffusive equilibrium between the sample and environment.\cite{daganResistivityPRL} this allowed us to carefully vary the oxygen concentration in the sample. The thickness of the film was measured using scanning electron microscope image of the cross-section. Resistivity and Hall were measured in a Van-der Pauw configuration using Keithley delta mode in a He3 refrigerator and a Lakeshore 370 AC resistivity bridge in the dilution refrigerator. Since the resistivity anisotropy factor $\rho_c/\rho_{ab}$ for PCCO is of the order of 10,000,\cite{Yucaxis} we convert the resistivity to sheet resistance using the CuO$_2$ planes spacing of $d=6\AA$ $(R_\square=\frac{\rho}{d})$.
\par
To ensure the absence of electron heating effects at the base temperature (30mK) we applied a magnetic field sufficient to partially quench superconductivity. For this field $\frac{d\rho}{dT}$ becomes very large. Consequently the temperature sensitivity is significantly enhanced compared to the nonsuperconducting state (higher magnetic fields). We used a current range small enough such that the resistivity is current independent. The resistivity data was symmetric for positive and negative magnetic fields, and independent of the field sweeping rate, therefore excluding eddy current heating effects. The typical sample resistance is of the order of a few Ohms at low temperatures such that $eV<k_BT$ even for base temperature, with $V$ being the voltage drop across the sample and the ohmic contacts.

\begin{figure}
\resizebox{1\columnwidth}{!}{%
  \includegraphics{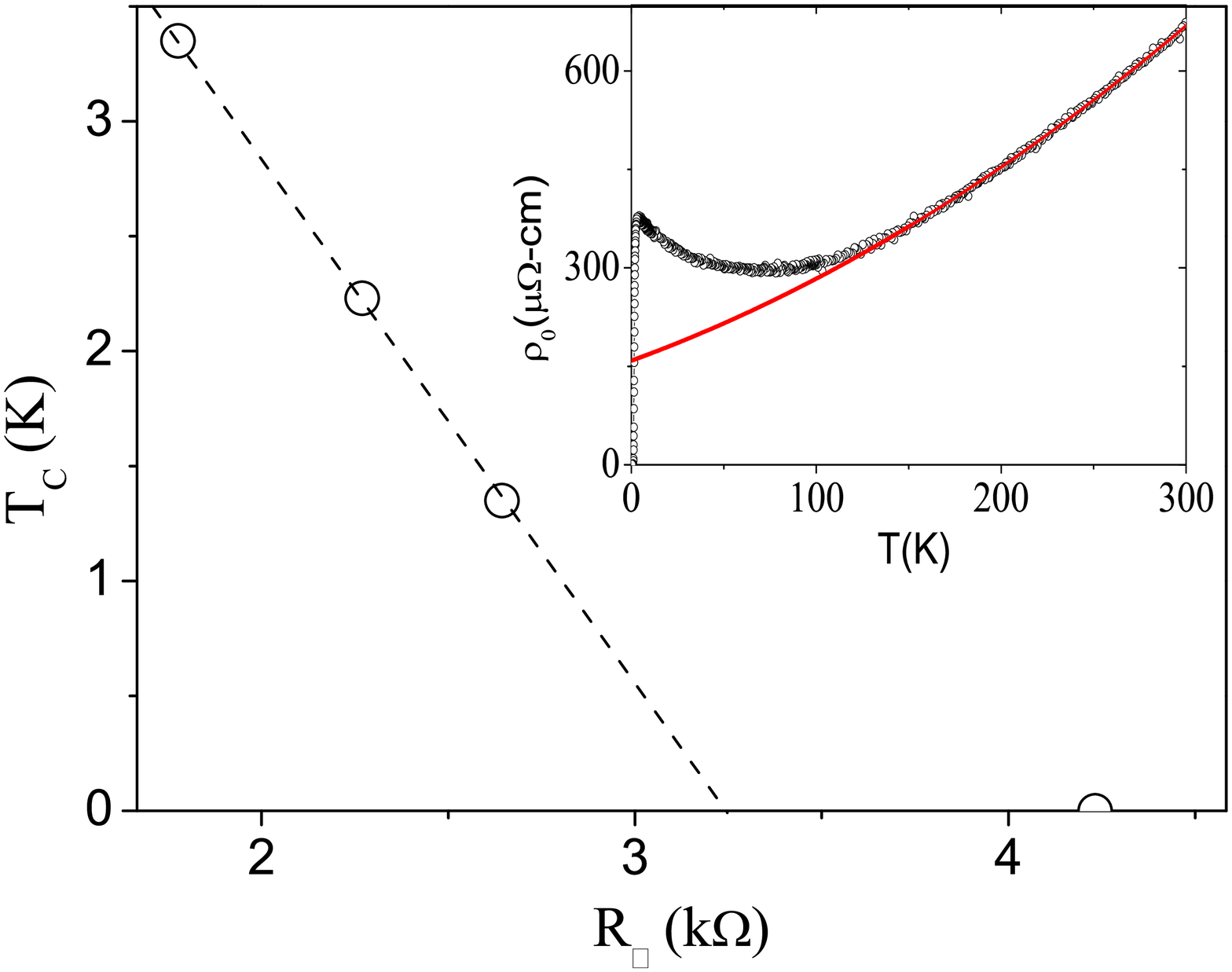}
}
\caption {(color on-line) Inset: Black circles are the resistivity data as a function of temperature for sample S2. The solid red line is a second degree polynomial fit to the high temperature regime (above the upturn). Main plot: The transition temperature $T_c$ for the various samples is plotted against $R_\square(0)$ obtained from an extrapolation to zero temperature of the fit, as demonstrated in the inset.
\label{residualres}}
\end{figure}

\par
\section {results}
In Fig.\ref{allrt} we show the sheet resistance as a function of temperature for the various oxygen contents. Sample S1 is not superconducting down to 20mK. The superconducting transition for S2-S4 is relatively sharp, indicative of homogenous oxygen (and cerium) content in these samples. It is interesting to note that superconductivity onsets at around a sheet resistance of 6.5K$\Omega$ close to the quantum resistance of Cooper pairs. This is expected for a two dimensional material undergoing a superconducting to insulating transition.\cite{FisherScalin}
Above the upturn the resistivity can be described by $\rho(T)=\rho_0+A(T)$ where $\rho_0$ is the residual resistivity. $A(T)$ changes very little from one sample to another, but $\rho_0$ increases with oxygen content. At this temperature regime we fit the resistivity data to $\rho(T)=\rho_0+aT+bT^2$. (see inset of Fig.\ref{residualres}) We extrapolate these fits to zero to obtain the zero temperature resistance $R_\square(0)=\rho_0/d$ for the various samples. We find that $T_c$ decreases linearly with $R_\square(0)$ (Fig.\ref{residualres}). This suggests a superconducting order parameter wich changes sign around the Fermi surface. Similar trends can be seen for oxygenated \PCCO~ crystals \cite{BrinkmannPhysicaC} and for irradiated \NCCO~ \cite{Woods} \PCCO~ \cite{higgins:104510} and YBa$_2$Cu$_3$O$_{7-\delta}$.\cite{Rullier}
\par
\begin{figure}
\resizebox{1.05\columnwidth}{!}{%
  \includegraphics{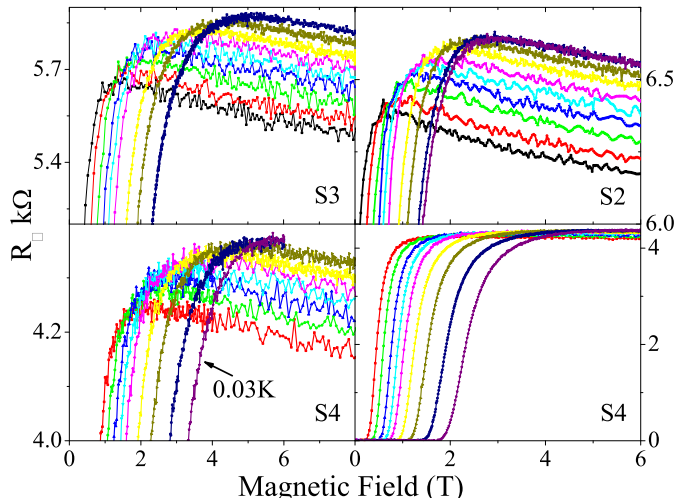}
}
\caption {The sheet resistance (samples S2-S4) as a function of magnetic field for various temperatures (from right to left): 2, 1.5, 1, 0.7, 0.5, 0.35, 0.22, 0.12, 0.05, and 0.03 K (0.03 K is not available for S3). Lower right panel: A broader field and resistance view.\label{allRH}}
\end{figure}

Fig.\ref{allRH} presents the field scan isotherms for the various superconducting samples. Starting with zero resistance at low fields the resistance increases rather steeply, indicative of a narrow flux flow regime. The rise in resistance is followed by a peak and then a negative magnetoresistance appears. We note that there is no clear crossing point for the isotherms even for sample S2 whose maximum sheet resistance is approximately the quantum resistance for Cooper pairs. \textit{The absence of a crossing point for all our samples casts doubt on the interpretation of the data in terms of a field tuned superconducting to insulating transition.}
\par
\begin{figure}
\resizebox{1.1\columnwidth}{!}{%
  \includegraphics{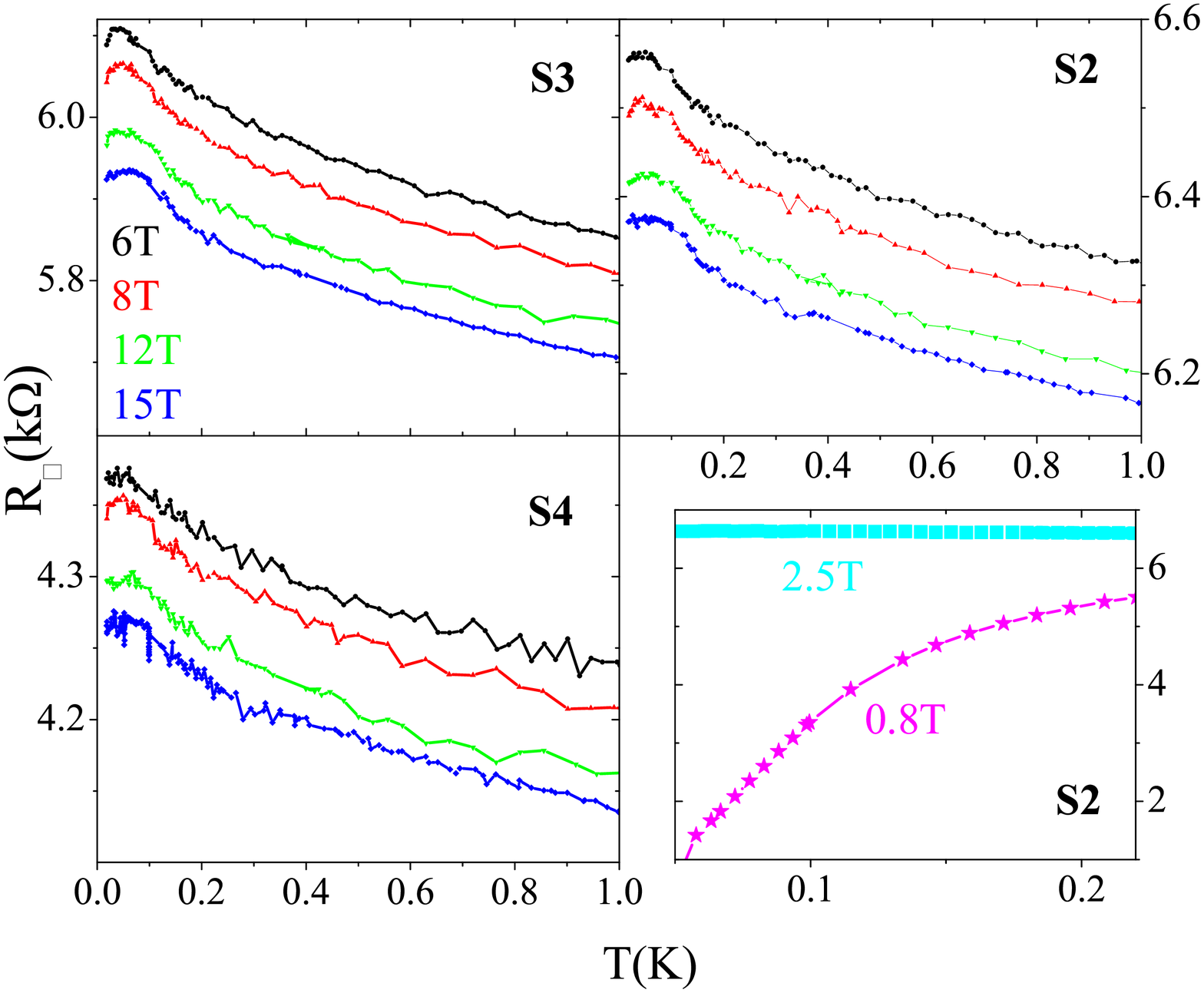}
}
\caption {The sheet resistance as a function of temperature for various magnetic fields and for the different samples. In the lower right panel we zoom on the low temperature and low field behavior of sample 2.\label{allRTvH}}
\end{figure}

The sheet resistance as a function of temperature for various magnetic fields for the superconducting samples: S2-S4 is shown in Fig.\ref{allRTvH}. As the temperature is lowered, the resistance increases. However, below approximately 0.1 K it reaches a saturation regime (where $dR/dT$ goes to zero). This saturation regime becomes slightly broader as the field is increased.In the bottom right panel Fig.\ref{allRTvH} we focus on the low field behavior of sample 2. While at high fields the resistance is constant below $~$0.1 K, at 0.8 T this sample exhibits a significant temperature dependence. In addition $eV<k_BT$ with e the electron charge, and $k_B$ the Boltzmann constant. Finally, looking at figures Fig.\ref{allRH} one notes that while at low fields the 30mK and 50mK resistance isotherms are clearly distinguishable, they merge at higher fields where the saturation in temperature is seen. This matching is not related to high resistance (resulting in higher voltages and higher electron heating) since for S4 the curves merge at 4.4 K$\Omega$ while for S2 they match at 6.5 K$\Omega$. We can therefore safely state that heating is not at the origin of the saturation.

\par
\section {Discussion}
Among our superconducting samples, sample 2 should theoretically exhibit features of two-dimensional SIT. It has the lowest $T_c$ and largest sheet resistance. The normal state sheet resistance at base temperature is very close to the critical resistance predicted for a two-dimensional SIT. It is therefore tempting to analyze the data in the context of a magnetic field tuned SIT for a 2D superconductor. Indeed, as the magnetic field is increased above a certain value there is a transition from a superconducting phase to another phase in which the sheet resistance increases when lowering the temperature. However, this insulating-like behavior is rather weak (Fig.\ref{allRTvH}), and the resistance saturates at low temperatures. This saturation does not completely rule out the SIT interpretation. It has been also observed for MoGe films,\cite{YazdaniKapitulnik} and was interpreted by Mason \etal\cite{MasonKapitulnik} as a consequence of coupling to a dissipative environment, presumably a background of delocalized fermions.
\par
At finite temperatures, the underlining quantum phase transition should manifests itself in the scaling behavior of the sheet resistance.\cite{FisherScalin} Therefore, the sheet resistance in both the superconducting and insulating phases can be described by a single universal scaling function: $R_\square(\delta,T)=R_cf(\delta/T^{\frac{1}{z\nu}})$, with $\delta=|B-B_c|$, $z$ and $\nu$ are the critical exponents $R_c$ is the resistance at the critical field $B_c$. We could not scale our data using $B_c=1.25 T$ and $R_c=6.5 k\Omega$ in either the strong \cite{SteinerPRB} or the weak disorder regimes with $z\nu=2.33$ and  $z\nu=1.33$ respectively. The absence of a well defined crossing point of the MR isotherms, the sheet resistance saturation and the failure of the scaling analysis suggest that a field tuned SIT cannot explain our data.

\begin{figure}
\resizebox{0.95\columnwidth}{!}{%
  \includegraphics{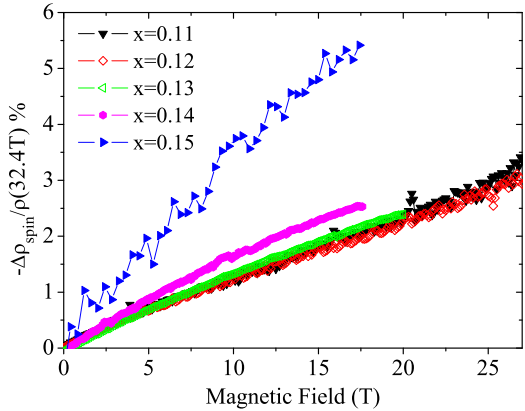}
}
\caption{(color online) Field dependence of the spin \MR~  at T=1.5 K (using the data of Ref.\cite{daganMR}. The spin \MR~ exhibits linear field dependence even at relatively high fields. The spin \MR~
drops to zero for $x\geq0.16$ at 1.5 K. \label{HdepspinMR}}
\end{figure}

Seikitani \etal\cite{seikitani} suggested that the resistivity upturn and the negative MR arise from Kondo scattering off Cu$^{2+}$ spins resulting from residual apical oxygen. However, it has been later shown \cite{daganMR} that there are two contribution to the MR: spin-MR and an orbital one. The spin-MR suddenly disappears for $x > 0.16$. This is the doping at which the resistivity upturn also disappears. This gives us a strong indication that these two quantities are closely related.
We analyze the data from Ref.\cite{daganMR} to extract the field dependence of the spin MR for the underdoped regime. To approximate the spin \MR~ for the superconducting samples we assume that when we apply a field
$\put(2,10){\vector(1,0){8}}H'=(H'_c,H'_{ab})$; the orbital \MR~
at that field can be subtracted out by measuring $\rho(H'_c,0)$.
Hence, the spin \MR~ of a field with an amplitude of
$(H=|\put(2,10){\vector(1,0){8}}H'|-H'_c)$ is:
$\Delta\rho_{spin}(H)=\rho(H'_c,0)-\rho(\put(2,10){\vector(1,0){8}}H')$.
The spin \MR~ obtained from this procedure is
plotted for various doping levels in figure \ref{HdepspinMR}
at 1.5 K. What we find is that a spin \MR~ exists for $x<0.16$ and the
field dependence of this \MR~ is linear at $H>2T$. This rules out the Cu$^{2+}$ Kondo scattering as a reason for the
spin \MR~ since in that case the \MR~ is expected to have a logarithmic
dependence on the magnetic field.
It is therefore reasonable to
relate the spin \MR~ to the antiferromagnetic order existing at these
doping levels. Scenarios such as scattering off
magnetic droplets have been proposed.\cite{daganMR} A recent theoretical work\cite{Hirschfeld_upturns} may point the way to a complete explanation of this problem for both electron and hole doped cuprates.
\par
In summary, we measured the resistivity as a function of temperature and magnetic field for a series of Pr$_{1.88}$Ce$_{0.12}$CuO$_{4-\delta}$ films with various oxygen concentrations. The critical temperature, T$_c$ is proportional to the residual resistivity as expected for $d$-wave superconductors. The \MR~ isotherms do not exhibit a clear crossing point as expected for a field-tuned superconductor-to-insulator (SIT) transition. A scaling behavior of the \MR~ predicted for SIT is not observed as well. We are therefore lead to the conclusion that a different mechanism governs the low temperature resistivity. In view of the proximity to an antiferromagnetic state and since a linear in field spin-\MR~ appears at the same doping level as the resistivity upturn, it is possible that magnetic droplets are at the origin of the enhanced scattering at low temperatures. This scenario has been previously suggested both experimentally\cite{daganMR} and theoretically.\cite{Hirschfeld_upturns} Since the low temperature upturn in resistivity is common to both hole and electron-doped cuprates it is possible that the magnetic droplets scenario describes the cuprates on both sides of the phase diagram.

\begin{acknowledgments}
This research was partially supported by the Bi-national science foundation under grant 2006385, the ISF under grant 1421/08 and the NSF under grant DMR-0653539. A portion of this work was performed at the National High Magnetic Field
Laboratory, which is supported by NSF Cooperative Agreement No.
DMR-0654118, by the State of Florida, and by the DOE. \end{acknowledgments}

\bibliographystyle{apsrev}
\bibliography{longtransport}
\end{document}